# Investigation of variable temperature Mössbauer spectrum of $YFe_{0.5}Cr_{0.5}O_3$ perovskite

Jingzhi Liu, Kai Wang, Lebin Liu, Jiajun Mo, Shiyu Xu, Haiqi Yang, Min Liu[*]

[*]e-mail liuhart@126.com

College of Nuclear Science and Technology, University of South China, Hengyang 421001, China.

**Abstract**—In this paper, we reported the preparation of $YFe_{0.5}Cr_{0.5}O_3$ by the sol-gel method and studied its structure and Mössbauer spectrum at variable temperatures. X-ray diffraction(XRD) analysis exhibits that the sample has the orthorhombic structure with the *Pnma* space group, and the energy dispersive spectroscopy (EDS) analysis shows that the sample has Fe/Cr = 1:1, indicating that the sample is Fe half-doped $YCrO_3$. The hyperfine parameters of the Mössbauer spectrum at room temperature confirm that the characteristics of $^{57}Fe$ in the sample were trivalent hexacoordinated high-spin(s=5/2), and the coexistence of doublet and the sextets at 250K indicate that the sample has superparamagnetic relaxation. The Mössbauer spectrum records the magnetic phase transition in the temperature range of 250K-300K.

## INTRODUCTION

Magnetic material is a kind of functional material. Through the study of the interaction and coupling between its spin, charge, lattice and electron degrees of freedom, soft magnetic materials, giant magnetic (magnetic storage) materials, magnetic recording materials, spin magnetic materials, semi-hard magnetic materials, and magnetoelectronic materials are gradually developed, which are widely used in the fields of force, heat, light, and electricity [1-10].

$ABO_3$-type rare earth ferrites ($T_N$~620-740K) and orthochromate ($T_N$~120-250K) have been widely studied due to their interesting properties, such as magnetization reversal, exchange bias and magneto-dielectric effect, especially in the field of ferromagnetism, ferroelectricity and ferroelasticity[11-13]. In the development of spintronics and new electronic devices, it is of great significance to develop the new multiferroic materials with internal coupling between spin and lattice degrees of freedom.

Researchers generally believe that multiferroic materials with strong magnetoelectric (ME) coupling at room temperature are the most promising data storage materials. However, there are two major problems in practical applications of the multiferroic materials, first of all, the ME coupling strength of $BiFeO_3$ does not meet the requirements of production applications at room temperature, secondly, the ordering temperature of perovskite with strong ME coupling deviates is far from room temperature[14,15]. However, through the physical study of the origin of ferroelectric and magnetic sequences in perovskite, Dzyaloshinsky-Moriya (DM) interaction, magnetoelectric coupling and mutual control, four research directions (beyond the $d^0$-$d^n$ scenario, A-site driven ferroelectric multiferroics, B-site driven ferroelectric multiferroics and double complex perovskite) were proposed by researchers to develop materials with strong ME coupling at room temperature[16,17]. In addition, people are also exploring the multifunctional property and structural flexibility of perovskite materials of pure inorganic compounds by organism substitution, which is called hybrid organic-inorganic new perovskite ($ABX_3$ type, A and X sites are replaced by organic cations and organic linkers, respectively)[18].

In recent decades, there are many reports of B site doping, and many discoveries have been made, and the study of $RFe_xCr_{1-x}O_3$ perovskite is a noteworthy part of it. For example, the Gitanjali Kolhatkar team improved the multiferroic properties of $BiFeO_3$ materials (polar and magnetic ordering at room temperature, which is very suitable for nonvolatile semiconductor memory) by chromium doping[15]. Below the magnetic ordering temperature, the ferroelectricity and antiferromagnetism of $DyFe_{0.5}Cr_{0.5}O_3$($T_N$=261K) coexist and show multiferroicity, and it has application promise in spintronics[19]. In 2005, the Azad team verified the existence of the magnetization reversal effect (MR) of $LaFe_{0.5}Cr_{0.5}O_3$ compounds, showing its application potential in the field of magnetocaloric devices[20].

There are many studies on the magnetic and dielectric properties of $YFe_{0.5}Cr_{0.5}O_3$ (YFC), but there are few studies on the variable temperature Mössbauer spectrum of YFC. In this paper, the nanometer sample powder of YFC was prepared by the sol-gel method. The related hyperfine structure of YFC and magnetic properties at different temperatures were given by variable temperature Mössbauer spectrum analysis.

## EXPERIMENTAL

### Synthesis of the $YFe_{0.5}Cr_{0.5}O_3$ perovskite

YFC was synthesized by the improved sol-gel method[21,22]. $Fe(NO_3)_3·3H_2O$, $Cr(NO_3)_3·9H_2O$ and citric acid were dissolved in deionized water in the ratio of n[Y2O3(99.9%)]: n[$Fe(NO_3)_3·3H_2O$ (analytical grade)]: n[$Cr(NO_3)_3·9H_2O$(analytic grade)]=2: 1: 1, n(transition mental):n(citric acid)=1: 2. Stir the solution until it is completely dissolved Then mix them together. The salt solution was slowly evaporated at 100 °C and stirred properly with a magnetic agitator to form an organic gel with uniform cation distribution. The gel was dried at 150 °C to form dendritic compounds, which were crushed into fine precursors after drying. The gel was decomposed at 600 °C for 12 h and then treated at 1100 °C for 12 h.

### Measurements

Using a Siemens D500 Cu-K diffractometer, diffraction angle data were collected at a speed of 0.05 °/s in a wide Bragg angle range (20°≤2θ≤80°). The crystal structure of the compound was studied using MDI-JADE software; The sample was analyzed by energy dispersive spectroscopy

(EDS). The Seeco W304 Musbourg spectrometer ($^{57}$Co/ Rh source) recorded the Musbourg spectrums of the sample at 12K, 200K, 225K, 250K, and RT(300K).

## RESULTS AND DISCUSSION
### XRD Analysis

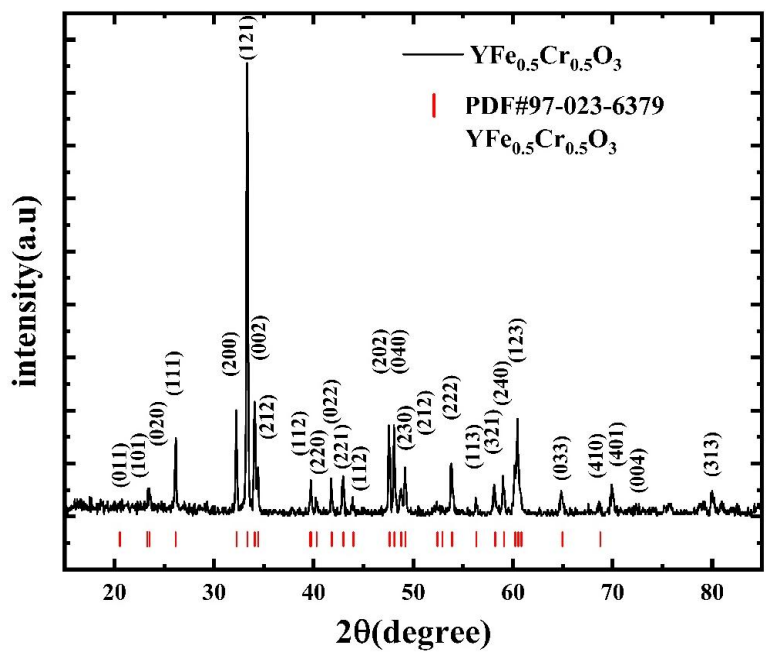

**Fig. 1.** XRD patterns of YFC

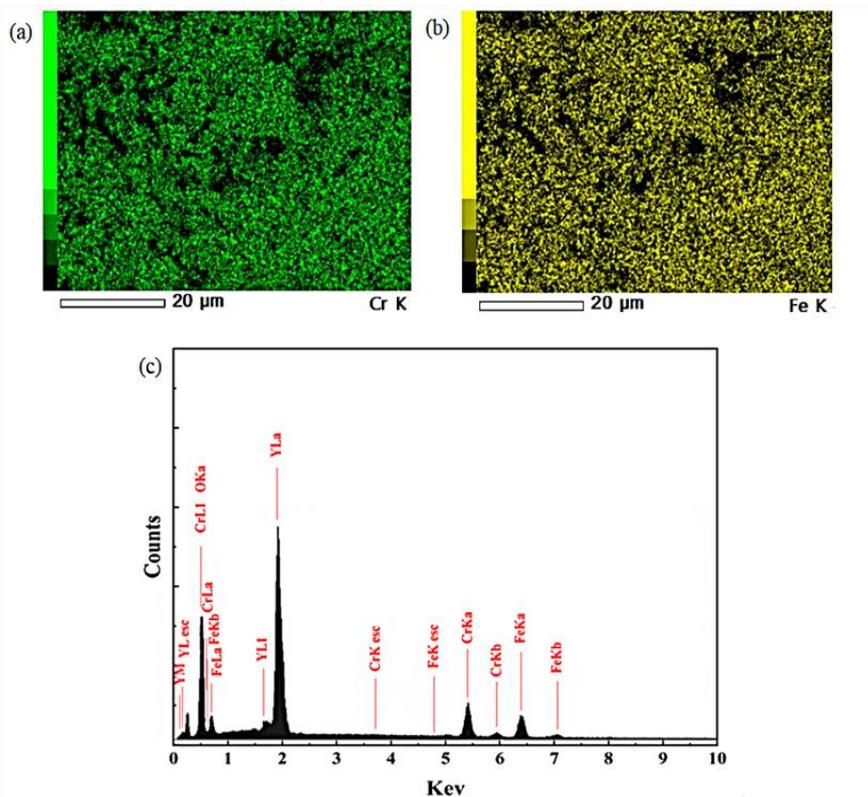

**Fig. 2.** EDS spectra for compositions of YFC

**Table 1.** The content of the Fe and Cr in YFC

| Element | Mass, % | Atom, % |
|---------|---------|---------|
| Fe K    | 11.30   | 11.99   |
| Cr K    | 11.08   | 12.62   |

The room temperature XRD pattern of the sample is shown in Fig. 1 By comparing the diffraction peaks of the sample with the JCPDS card (YFe$_{0.5}$Cr$_{0.5}$O$_3$, # 97-023-6379, it is found that all the diffraction peaks belong to the Pnma structure with Orthorhombic perovskite structure, which is consistent with previous studies on YFC[23]. The lattice parameters, the volume of the unit cell, and the average grain size of YFC were calculated by the Bragg formula and Scherrer formula, a=5.547 Å; b=7.559 Å; c=5.256 Å; v=220.21 Å$^3$; D=62.8nm. This is in line with the previous research[23]. Meanwhile, the crystal cell parameters of YCrO$_3$ are a=7.610 Å, b=7.540 Å, and c = 7.610 Å [24]. It is seen that the lattice parameters increase after iron ions replace chromium. In addition, the sample was confirmed to be Fe half-doped YCrO$_3$ by EDS analysis (Fig. 2 and Table 2 ).

*Mössbauer Spectroscopy Analysis*

The temperature-variable Mössbauer spectrum of YFC is shown in Fig. 2, and the Mössbauer spectrum at room temperature has two lines (a doublet and singlet). It can be seen that the sample is in a paramagnetic state and there is no impurity phase in the sample. The phenomenon is the same as that of Roberto Salazar-Rodriguez's YFe$_{0.5}$Cr$_{0.5}$O$_3$ room temperature Mössbauer spectrum study[25]. When the temperature decreases to 250K, the singlet disappears, the area of the doublet decreases, and three sextets appear. When the temperature continues to decrease to 225K, the area of the doublet decreases, and the area of the sextet increases. The same phenomenon occurs when the temperature drops to 200K. When the temperature decreases to 12K the doublet disappears as well as the superparamagnetism of the sample disappears.

For Mössbauer nuclides, the quadruple splitting is determined by the electric field gradient caused by the asymmetric distribution of charge around the nucleus, while isomer shift is affected by the s electron density around the nucleus, and has a certain relationship with the coordination number. The chemical bond properties, electron distribution and spin states of Mössbauer nuclides can be reflected by isomer shift. Table 1. shows the Mössbauer hyperfine parameters of the sample. (isomer shift, quadruple splitting/shift, line width and hyperfine field).

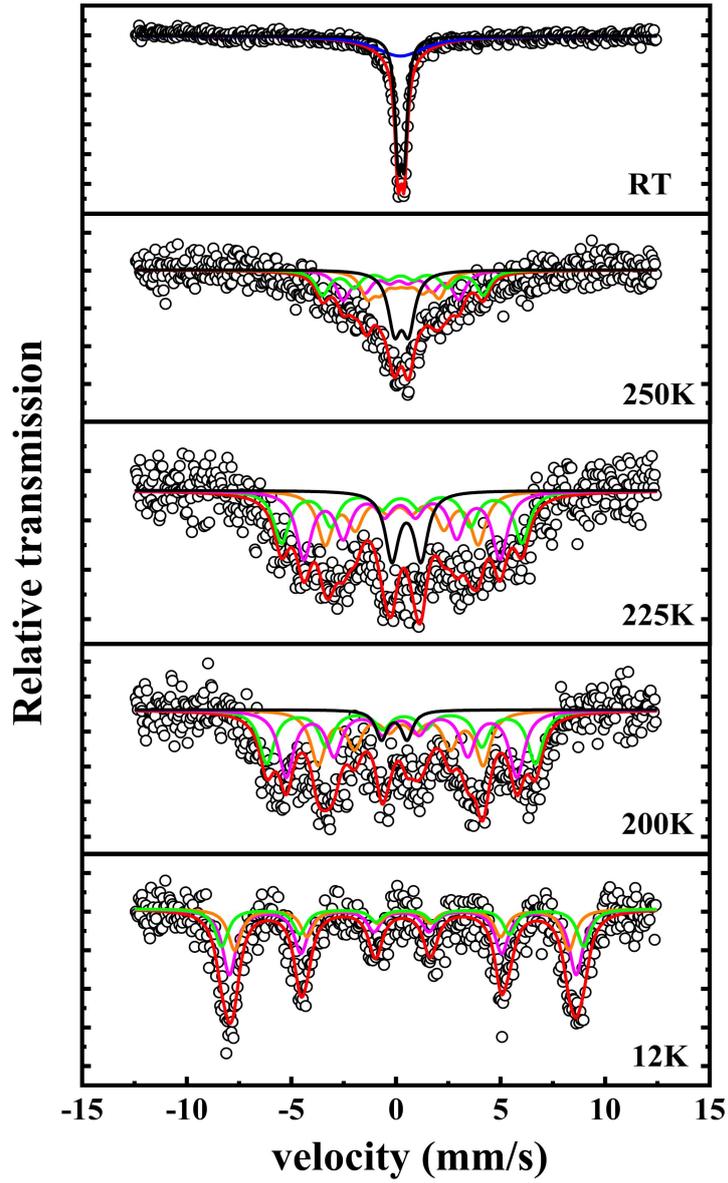

**Fig. 3.** $^{57}$Fe Mössbauer spectra of YFC

**Table 2.** Mössbauer parameters of YFC at various temperature

| Temperature | | H, T | QS, mm/s | IS, mm/s | Area, % | Γ, mm/s |
|---|---|---|---|---|---|---|
| RT | Singlet | —— | —— | 0.200 | 40.8 | 1.582 |
| | Doublet | —— | 0.295 | 0.239 | 59.2 | 0.354 |
| 250K | Sextet1 | 17.20 | 0.100 | 0.179 | 27.1 | 0.737 |
| | Sextet2 | 10.68 | 0.099 | 0.294 | 23.8 | 0.772 |
| | Sextet3 | 23.68 | 0.101 | 0.269 | 23.8 | 0.844 |
| | Doublet | —— | 0.668 | 0.257 | 25.3 | 0.582 |
| 225K | Sextet1 | 22.73 | -0.100 | 0.221 | 25.8 | 0.882 |
| | Sextet2 | 35.72 | 0.055 | 0.248 | 25.8 | 0.882 |
| | Sextet3 | 29.12 | 0.092 | 0.237 | 33.4 | 0.882 |
| | Doublet | —— | 1.058 | 0.500 | 15.0 | 0.782 |
| 200K | Sextet1 | 24.54 | -0.082 | 0.269 | 28.9 | 0.866 |
| | Sextet2 | 39.89 | -0.143 | 0.321 | 29.0 | 0.873 |
| | Sextet3 | 34.28 | 0.303 | 0.256 | 36.1 | 0.871 |
| | Doublet | —— | 1.221 | -0.078 | 6.0 | 0.660 |
| 12K | Sextet1 | 51.51 | 0.069 | 0.294 | 46.5 | 0.791 |
| | Sextet2 | 49.65 | 0.100 | 0.332 | 28.0 | 0.739 |
| | Sextet3 | 53.78 | -0.100 | 0.375 | 25.5 | 0.756 |

Isomer Shift (IS), Quadruple Splitting/Shift (QS), Lorentzian Linewidth (Γ), Hyperfine Magnetic Field(H)

The IS and QS values of YFC at room temperature are 0.239mm/s and 0.295mm/s, respectively, which is similar to IS=0.231mm/s and QS=0.290mm/s reported by Roberto Salazar-Rodriguez[25]. Combined with the previous studies on $RFe_{0.5}Cr_{0.5}O_3$(R=Gd, Sm) nanoparticles at room temperature[21,22], the IS value and weak QS value at room temperature indicate that the $^{57}Fe$ in the sample is a trivalent hexacoordinated high-spin(s=5/2) iron ion. Previous studies have shown that the distribution of the hyperfine field produced by $^{57}Fe$ depends on the number of $Cr^{3+}$ ions in the nearest neighbor sites of $Fe^{3+}$ ion[26]. The magnetization of the Fe atom is affected by nearest neighbor cations through exchanging interactions, The hyperfine field decreases with the increase of the nearest neighbor $Cr^{3+}$ ion of $Fe^{3+}$ ion, In the 12K~250K temperature range, when the temperature is constant, the area of the sextet located in the middle value of the hyperfine field is the largest. Meanwhile, The area of the largest sextet of the hyperfine field is equal to that of the smallest sextet of the hyperfine field. These phenomenons can also verify the half-doping of Fe at the B site (b-site Fe/Cr=1:1). when the sample is in a magnetic order state, the magnetoelectric combined hyperfine interaction will lead to the complex change of QS. In general, The larger specific surface area and chemical inhomogeneity caused by the local environment and nano-size will lead to larger linewidth[27], indicating that the nano-crystallinity of the sample maintains well in the whole test temperature range.

The appearance of singlet at room temperature and sextets at 250K indicates that the magnetic phase transition occurs in this temperature range, and the appearance of the doublet is attributed to the superparamagnetism of the sample. This is different from the conclusion of Salazar-Rodriguez, which uses the combustion method to prepare the sample, which leads to a too large grain size (bigger than 100 nm). The appearance of superparamagnetism is generally related to the particle size and temperature of the sample. When the blocking temperature $T_B$ corresponding to a certain size is lower than the field temperature, the sample shows superparamagnetism[28]. The blocking temperature($T_B$) is given by the following equation, $T_B \approx \frac{KV}{25k_B}$, where $K$ is the magnetic anisotropy constant, V is the sample size, and $k_B$ is the Boltzmann constant. With the decrease of the external field temperature, the number of samples whose blocking temperature is lower than the external field temperature decreases, resulting in the decrease of the doublet area. The coexistence of doublet and sextets in the range of 200K~250K and the disappearance of doublet at 12K indicate that the sample particle size has a considerable distribution[27]. In the range of 12K~250K, there are three sextets in the Mossbauer spectrum of the sample, which may be related to the arrangement of three atoms in the YFC cell ($Fe^{3+}$ and $Cr^{3+}$ ions are in an octahedron surrounded by oxygen)[25].

## CONCLUSIONS

Nano-sample YFC was prepared by the sol-gel method. The X-ray diffraction results of the samples show the orthorhombic perovskite structure (pnma space group). The grain size calculated by the Debye-Scheler formula is about 62.8nm. The Mössbauer spectrum parameters of the samples at room temperature show the characteristics of $^{57}Fe$ trivalent six-coordinated high spin (s=5/2). With the decrease in temperature, the secondary Doppler energy shift leads to the decrease of QS, and the larger linewidth also indicates that the nano-crystallinity of the sample is well maintained in the whole measured temperature range. The coexistence of doublet and sextets of Mossbauer spectrum in the 200K-250K range indicates that the sample has superparamagnetic relaxation. With the decrease in temperature, the magnetic phase transition occurred in YFC, the area of doublets decrease and the area of sextets increased.

## CONFLICT OF INTEREST

The authors declare that they have no conflicts of interest.

## FUND

The research was supported partly by National Natural Science Foundation of China (grant number 12105137), the Natural Science Foundation of Hunan Province, China (grant number 2020JJ4517), Research Foundation of Education Bureau of Hunan Province, China (grant number 19C1621,19A434),the National Undergraduate Innovation and Entrepreneurship Training Program Support Projects of China (Grant No. 20200112, 202110555026).

## DATA AVAILABILITY

All data used during the study appear in the submitted article

## AUTHORS' CONTRIBUTIONS

Jingzhi Liu and Min Liu conceived and designed the experiments. Lebin Liu, Jingzhi Liu and Kai Wang carried out the XRD and Mössbauer Spectroscopy experiments. JingZhi Liu,Jiajun Mo and Shiyu Xu analyzed the data. Jingzhi Liu,Lebin Liu, Jiajun Mo, ShiYu Xu,HaiQi Yang and Kai Wang wrote the paper. All authors discussed the results and contributed to the paper.

TABLES

**Table 1.** The content of the Fe and Cr in YFC

| Element | Mass, % | Atom, % |
|---------|---------|---------|
| Fe K | 11.30 | 11.99 |
| Cr K | 11.08 | 12.62 |

**Table 2.** Mössbauer parameters of YFC at various temperature

| Temperature | | H, T | QS, mm/s | IS, mm/s | Area, % | Γ, mm/s |
|---|---|---|---|---|---|---|
| RT | Singlet | —— | —— | 0.200 | 40.8 | 1.582 |
| | Doublet | —— | 0.295 | 0.239 | 59.2 | 0.354 |
| 250K | Sextet1 | 17.20 | 0.100 | 0.179 | 27.1 | 0.737 |
| | Sextet2 | 10.68 | 0.099 | 0.294 | 23.8 | 0.772 |
| | Sextet3 | 23.68 | 0.101 | 0.269 | 23.8 | 0.844 |
| | Doublet | —— | 0.668 | 0.257 | 25.3 | 0.582 |
| 225K | Sextet1 | 22.73 | -0.100 | 0.221 | 25.8 | 0.882 |
| | Sextet2 | 35.72 | 0.055 | 0.248 | 25.8 | 0.882 |
| | Sextet3 | 29.12 | 0.092 | 0.237 | 33.4 | 0.882 |
| | Doublet | —— | 1.058 | 0.500 | 15.0 | 0.782 |
| 200K | Sextet1 | 24.54 | -0.082 | 0.269 | 28.9 | 0.866 |
| | Sextet2 | 39.89 | -0.143 | 0.321 | 29.0 | 0.873 |
| | Sextet3 | 34.28 | 0.303 | 0.256 | 36.1 | 0.871 |
| | Doublet | —— | 1.221 | -0.078 | 6.0 | 0.660 |
| 12K | Sextet1 | 51.51 | 0.069 | 0.294 | 46.5 | 0.791 |
| | Sextet2 | 49.65 | 0.100 | 0.332 | 28.0 | 0.739 |
| | Sextet3 | 53.78 | -0.100 | 0.375 | 25.5 | 0.756 |

Isomer Shift (IS), Quadruple Splitting/Shift (QS), Lorentzian Linewidth (Γ), Hyperfine Magnetic Field(H)

FIGURE CAPTIONS

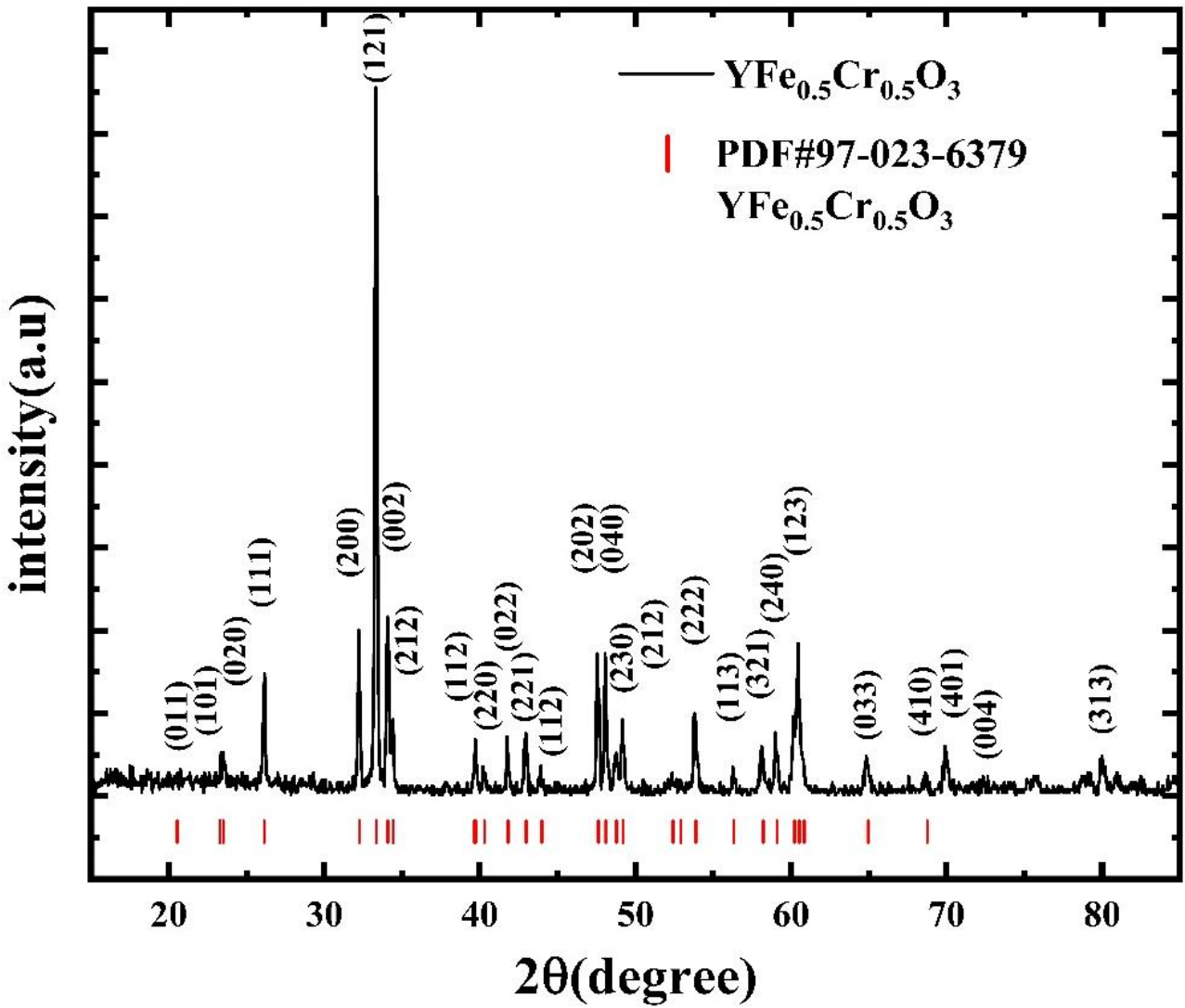

**Fig. 1.** XRD patterns of YFC

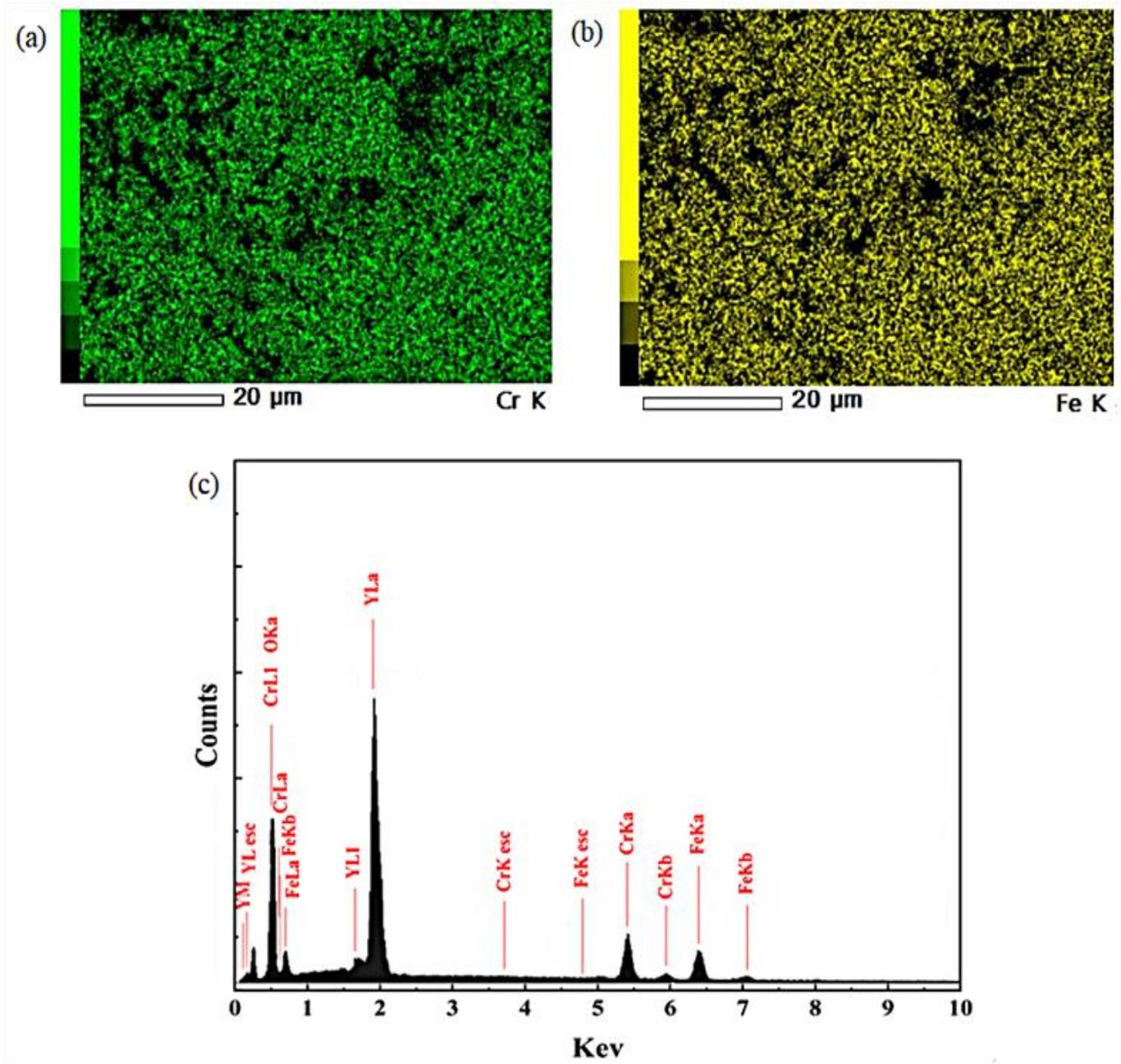

**Fig. 2.** EDS spectra for compositions of YFC

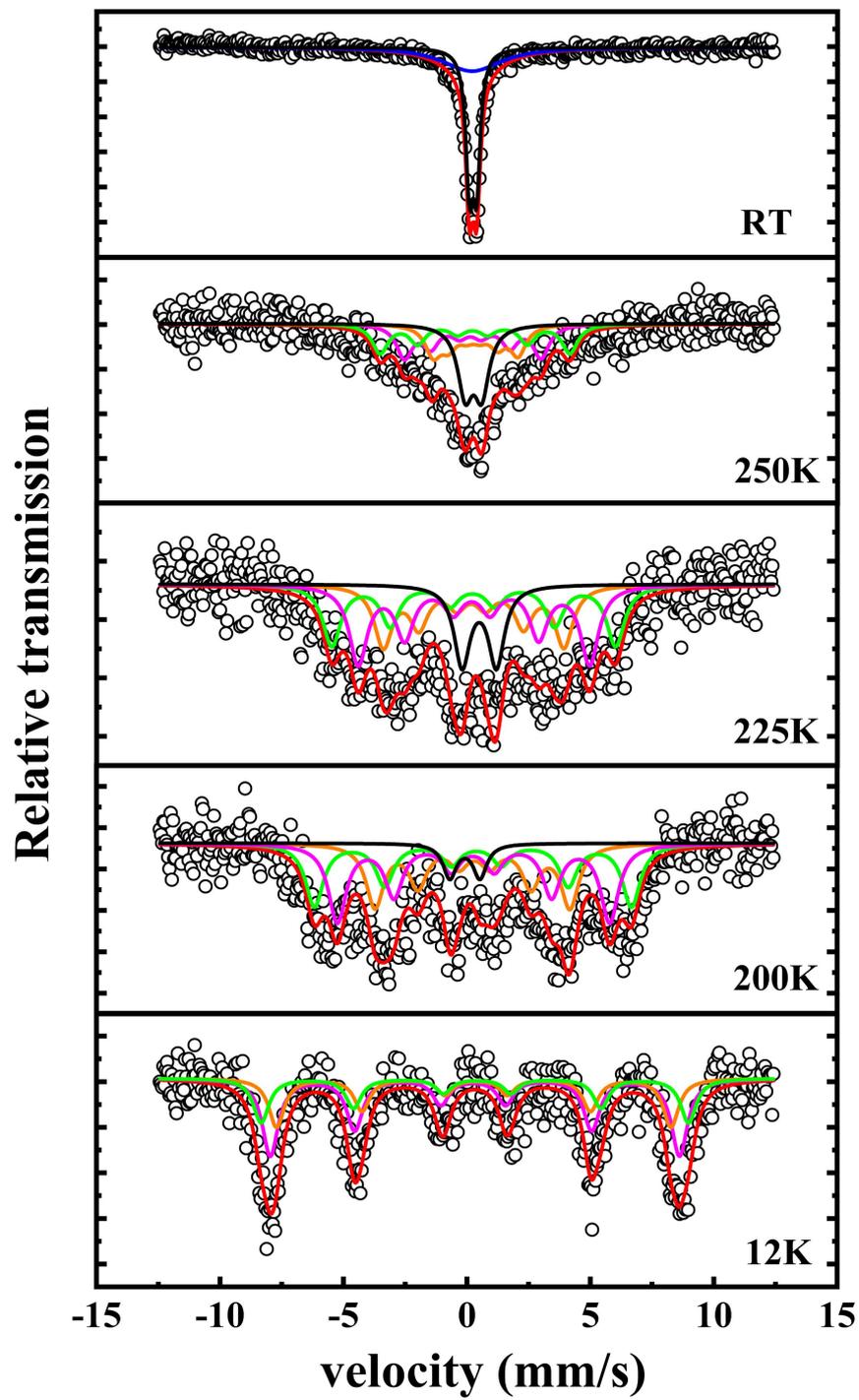

**Fig. 3.** $^{57}$Fe Mössbauer spectra of YFC